\documentclass[showpacs,twocolumn,aps,pre]{revtex4}

\def \be{\begin{equation}} %
\def \ee{\end{equation}} %
\def \im{ {i} } 
\def \dif { \mathrm{d} } 
\def \p {\phi}
\def \t {\theta}

\def \siteno{ {N} } 
\def\l{\langle}
\def\r{\rangle}

\def \vol {\bf}
\def \journal {}

\usepackage{graphicx}
\usepackage{amsmath}

\begin{document}

\title{Exact solution of close-packed dimers on the kagome lattice}

\author{Fa Wang$^{1,2}$ and F. Y. Wu$^3$}
\affiliation{
$^1$Department of Physics, University of California, Berkeley, California 94720 \\
$^2$Material Sciences Division, Lawrence Berkeley National Laboratory, Berkeley, California 94720 \\
$^3$Department of Physics, Northeastern University, Boston, Massachusetts 02115 }

\date{Printed \today}

\begin{abstract}
It is well-known that exact enumerations of close-packed dimers can be carried 
out for two-dimensional lattices. While details of results are now known for most 
lattices, due to the unique nature of the lattice structure,
there has been no complete analysis for the kagome lattice.
Here we derive the close-form expression $(1/3) \ln (4 x y z)$ 
for the free energy of close-packed dimers on the kagome lattice, 
where $x,y,z$ are dimer weights. We use two different approaches: 
the Kasteleyn method of evaluating a Pfaffian and an alternative 
vertex model formulation.  Both methods lead to the same final expression.
The correlation function between two dimers at a distance equal or greater than two
lattice spacings is found to vanish identically.
 \end{abstract}

\pacs{05.50.+q,04.20.Jb,02.10.Ox}

\maketitle

\section{Introduction}\label{sec:intro}

A central problem in statistical physics
is the enumeration of close-packed dimers on lattices. 
The origin of the problem has a long history tracing back to the 1937 paper 
of Fowler and Rushbrooke \cite{fowler:} in their attempt of enumerating the absorption 
of diatomic molecules on a surface. 
A milestone in the history of the dimer problem is the exact solution for the square lattice obtained
by Kasteleyn \cite{kasteleyn:physica} and Temperley and Fisher \cite{temperley:} in 1961. 
Indeed, the method of Kasteleyn is quite general and applicable 
to all planar lattices \cite{kasteleyn:jmathphys}. 
Exact results obtained in this way are summarized in a recent 
review \cite{fywu:review} for a number of two-dimensional lattices. 

In the case of the kagome lattice, however, 
there has been no complete analysis of the dimer problem other than
studies of pure dimer enumerations most of which are numerical and series expansions
(see \cite{phares:} and references cited therein).
In recent years there has been considerable interest
in the study of physical phenomena on the kagome lattice. 
These range from high-$T_c$ superconductivity \cite{obredors:}, 
Heisenberg antiferromagnets \cite{elser:, helton:, ofer:, mendels:, bert:}, 
quantum dimers \cite{misguich:}, 
to the occurrence of spin-liquid states \cite{wang:PRB}.
It has also been shown that the consideration of close-packed dimers
is related to the ground state of a quantum dimer model \cite{rokhsar-kivelson:}.
In light of these developments, it is of pertinent interest to take a fresh look 
at close-packed dimers for the kagome lattice.

In this paper we consider this problem, and derive the closed-form expression 
\be
f_{\mathrm{kagome}}(x,y,z)=(1/3)
\ln(4 x y z)
\label{equ:result1}
\ee
for the free energy (for definition of terms see below), a formula quoted in \cite{fywu:review}.
As exact solutions for other lattices are invariably of the form of
a double integral akin to the Onsager solution of the Ising model \cite{fywu:review},
the very simple expression of the solution (\ref{equ:result1}) and its logarithmic
dependence on dimer weights are
novel and unique. It points to the special role played by the kagome lattice 
(which often makes a problem more amenable),
and suggests that caution must be taken in generalizing physical
results derived from the kagome lattice.
For example, as we shall see in  Sec.\,\ref{sec:correlation} below,
the correlation between two dimers is identically zero 
at distances equal or greater than two lattice spacings on the kagome lattice, but this
conclusion does not hold for other lattices.
 
\begin{figure}
\includegraphics{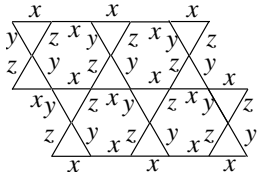}
\caption{The kagome lattice $\mathcal{L}$ with dimer weights $x,y,$ and $z$ in the three principal directions.}
\label{fig:kagome}
\end{figure}

The kagome lattice  is shown in Fig.~\ref{fig:kagome}, where $x,y,$ and $z$ are 
dimer weights along the three principal directions. 
We denote the lattice by $\mathcal{L}$. Let $\siteno$ ($=$ even) be 
the number of sites of $\mathcal{L}$, so the lattice can be completely covered by $\siteno/2$ 
dimers. The dimer generating function is defined to be the summation
\be
Z_{\mathrm{kagome}}(x,y,z)=\sum_{\mathrm{dimer\ coverings}}{ x^{n_x} y^{n_y} z^{n_z} }
\label{equ:defZ}
\ee
over all close-packed dimer configurations of $\mathcal{L}$. Here, $n_x$ is the number of 
dimers with weight $x$, etc., subject to $n_x+n_y+n_z=\siteno/2$. Our goal is to 
evaluate the {\it per-dimer} free energy
\be 
f_{\mathrm{kagome}}(x,y,z)=\lim_{\siteno\rightarrow \infty}{\frac{1}{\siteno/2}\ln Z_{\mathrm{kagome}}(x,y,z)}
\label{equ:deff}
\ee
in a close form. Past attempts  have been confined to 
enumerations of $f(1,1,1)$. Here we consider the problem for 
general $x,y,z$, a consideration which can find application on anisotropic kagome system 
such as the volborthite antiferromagnet \cite{bert:}. 

We derive the solution (\ref{equ:result1}) using two different methods: 
the Kasteleyn method of evaluating a Pfaffian and 
alternately a method of a vertex model formulation, 
which we describe in the next two sections. 

\section{Pfaffian approach}\label{sec:pfaffian}

The first step of the Pfaffian method is to find a Kasteleyn 
orientation \cite{kasteleyn:physica} of lattice edges.  
A Kasteleyn orientation of a planar lattice is an orientation of edges
such that every transition cycle consisting of a loop of edges derived 
from the superposition of two dimer coverings has an 
odd number of arrows pointing in the clockwise direction,
a property which we term as {\it clockwise-odd}. 
While Kasteleyn \cite{kasteleyn:jmathphys} has demonstrated that such an orientation is
possible for all planar graphs, the actual orientation of edges for a given lattice,
or graph, still needs to be  
worked out, and the crux of the matter of the
Kasteleyn method is the finding of the appropriate clockwise-odd orientation. 

\begin{figure}
\includegraphics{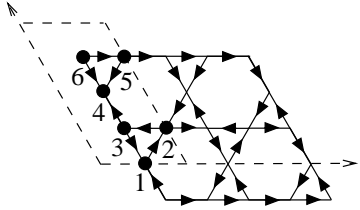}
\caption{The Kasteleyn orientation of the kagome lattice. A unit cell is the region bounded
by dashed lines containing the six sites numbered $1,\dots, 6$.}
\label{fig:KO}
\end{figure}

For the kagome lattice a Kasteleyn orientation can be taken as that
shown in Fig.~\ref{fig:KO}. The kagome lattice is composed
of up-pointing and down-pointing triangles. The orientation 
in Fig.~\ref{fig:KO} consists of orienting
all up-pointing triangles and every other down-pointing
triangles in the counter-clockwise direction, with the other half of the down-pointing
triangles oriented as shown.  In this orientation a unit cell of the lattice consists
of the 6 sites forming two neighboring down-pointing triangles,
which are numbered $1,\dots, 6$ as shown.
Our orientation is different from that used in \cite{phares:}.

To see that the orientation in Fig.~\ref{fig:KO} is indeed a Kasteleyn orientation, 
we note that all transition cycles in Fig.~\ref{fig:KO} 
are clockwise-odd.  As  all transition cycles on the lattice can
be formed by deforming those in Fig.~\ref{fig:KO} without altering the
clockwise-odd property, all  transition cycles are also clockwise-odd
so the orientation in Fig.~\ref{fig:KO} is a good Kasteleyn 
orientation \cite{boundary:}. 
This is essentially the argument of Kasteleyn \cite{kasteleyn:jmathphys}.

After rendering the Kasteleyn orientation, we can write the dimer partition function as 
a Pfaffian. As the analysis is now standard we follow the standard procedure 
[see, for example, Eq. (4.34) of \cite{montroll:book}] and arrive at the result
\begin{eqnarray}
& & f_{\rm kagome}(x,y,z)\nonumber\\
&=& \bigg( \frac 1 3 \bigg) \frac 1 {(2\pi)^2} 
 \int_0^{2\pi}\int_0^{2\pi}  \ln {\rm Pf}
[ M(\theta,\phi)]\, \dif \theta \, \dif \phi \, \nonumber\\
 &=&  \bigg( \frac 1 6 \bigg) \frac 1 {(2\pi)^2}  \int_0^{2\pi}\int_0^{2\pi}  \ln {\rm det}
| M(\theta,\phi)| \, 
   \dif \theta\, \dif \p \ .
\label{Equ:integral}
\end{eqnarray}
Here the factor $1/3$ in the first line comes from the fact that there are 3 dimers
per unit cell in a close-packed configuration, and
${\rm Pf} [M(\t,\p)] = \sqrt{ {\rm det} |M(\t,\p)| }$ is 
the Pfaffian of the  $6\times 6$ matrix
\begin{widetext}
\begin{eqnarray}
M(\t,\p) &=& a(0,0) +a(1,0) e^{\im\t} +a(-1,0)e^{-\im\t} +a(0,1) e^{\im\p} +a(0,-1)e^{-\im\p} + a(1,1) e^{\im(\t+\p)} +a(-1,-1)e^{-\im(\t+\p)}\nonumber \\
&=& 
\begin{pmatrix}
0 & z & -y & 0 & z e^{-\im(\t+\p)}  & -y e^{-\im \p}\cr
-z & 0 & x(1+e^{\im\t})  & -z e^{\im\t}&0&0\cr
y & -x(1+e^{-\im\t}) & 0 & y & 0 & 0\cr
0 & z e^{-\im\t} & -y & 0 & -z & -y \cr
-z e^{\im(\t+\p)} & 0 & 0 & z & 0 & x(-1+ e^{\im\t})\cr
y e^{\im\p} & 0 & 0 & y & x(1-e^{-\im\t}) & 0 \cr
\end{pmatrix}.
\label{Equ:matrix}
\end{eqnarray}
\end{widetext}
The $a$-matrices are  read off from Fig.~\ref{fig:KO} to be
\begin{subequations}
\begin{eqnarray}
a(0,0) &=& \begin{pmatrix} 0 & z & -y & 0 & 0 & 0 \cr
                  -z &0 & x & 0 & 0 & 0 \cr
                 y & -x & 0 & y & 0 & 0 \cr
                0 & 0 & -y & 0 & -z & -y \cr
                0 & 0 & 0 & z & 0 & -x \cr
               0 & 0 & 0 & y & x & 0 \cr \end{pmatrix}\\
a(-1,0) &=& - a^T(1,0)
\end{eqnarray}
\begin{eqnarray}
a(0,-1) &=& -a^T(0,1)\\
a(-1,-1) &=& - a^T(1,1)\\
a(1,0)  &=&\begin{pmatrix} 0 & 0 & 0 & 0 & 0 & 0 \cr
                  0 &0 & x & -z & 0 & 0 \cr
                 0 & 0 & 0 & 0 & 0 & 0 \cr
                0 & 0 & 0 & 0 & 0 & 0 \cr
                0 & 0 & 0 & 0 & 0 & x \cr
               0 & 0 & 0 & 0 & 0 & 0 \cr \end{pmatrix}
\end{eqnarray}
\begin{eqnarray}
a(0,1)  &=& \begin{pmatrix} 0 & 0 & 0 & 0 & 0 & 0 \cr
                  0 &0 & 0& 0 & 0 & 0 \cr
                 0 & 0 & 0 & 0 & 0 & 0 \cr
                0 & 0 & 0 & 0 & 0 & 0 \cr
                0 & 0 & 0 & 0 & 0 & 0 \cr
               y & 0 & 0 & 0 & 0 & 0 \cr \end{pmatrix}\\
a(1,1) &=& \begin{pmatrix}0 & 0 & 0 & 0 & 0 & 0 \cr
                  0 &0 & 0 & 0 & 0 & 0 \cr
                 0 & 0 & 0 & 0 & 0 & 0 \cr
                0 & 0 & 0 & 0 & 0 & 0 \cr
                -z & 0 & 0 & 0 & 0 & 0 \cr
               0 & 0 & 0 & 0 & 0 & 0 \cr \end{pmatrix}
\end{eqnarray}
\end{subequations}
where the superscript $T$ denotes the matrix transpose.

The evaluation of the determinant in Eq.~(\ref{Equ:integral}) gives the surprisingly simple result
\be
\det[M(\t,\p)]=16 x^2 y^2 z^2\ .\label{Equ:determinant}
\ee
The substitution of Eq.~(\ref{Equ:determinant}) into Eq.~(\ref{Equ:integral}) now
yields Eq.~(\ref{equ:result1}). The expression (\ref{Equ:determinant}) and result 
(\ref{equ:result1}) have previously been obtained
for $x=y=z=1$ in \cite{phares:} for pure dimer enumerations. 

\section{Vertex-model approach}\label{sec:vertex}
\begin{figure}
\includegraphics{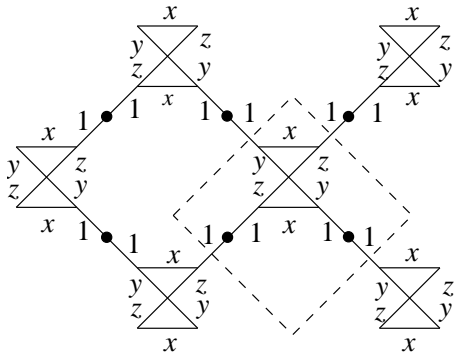}
\caption{An extended kagome lattice $\mathcal{L}'$ constructed 
by inserting a decorating site attached to two inserted edges of weight $1$ as shown.
The decorating sites are denoted by solid circles.
The unit cell is the region bounded by dashed lines. Repeating unit cells form a square lattice.}
\label{fig:extendedkagome}
\end{figure}

The kagome dimer problem can also be solved using a
vertex model approach which is conceptually simpler. 
This involves the mapping of the dimer problem into 
a vertex model for which the solution is known. 

The first step of the mapping is to introduce the extended kagome lattice 
$\mathcal{L}'$ of Fig.~\ref{fig:extendedkagome}. 
The extended lattice $\mathcal{L}'$ is constructed from $\mathcal{L}$ by introducing 
$4N/3$ extra lattice edges with
weight $1$ and $2N/3$ new (decorating) sites as shown. By inspection it is 
clear that a bijection exists between dimer configurations on $\mathcal{L}$ and $\mathcal{L}'$. 
This permits us to consider instead the dimer problem on $\mathcal{L}'$. 

The dimer problem on $\mathcal{L}'$ is next mapped onto a vertex model. 

\begin{figure}
\includegraphics{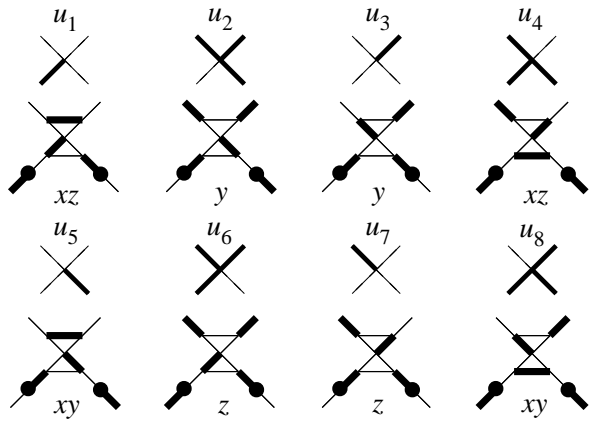}
\caption{Mapping between vertex and dimer configurations and the corresponding weights.}
\label{fig:vertex-dimer}
\end{figure}

The extended lattice $\mathcal{L}'$ consists of $\siteno/3$ unit cells each of which is 
the region bounded by dashed lines
shown in Fig.~\ref{fig:extendedkagome}. The unit cells form a square lattice $\mathcal{S}$. We next map dimer 
configurations on $\mathcal{L}'$ into vertex configurations on $\mathcal{S}$, by regarding the 4 edges 
extending from a unit cell of $\mathcal{L}'$ as the 4 edges incident to a site on $\mathcal{S}$. To each 
extending edge on $\mathcal{L}'$ covered by a dimer, draw a {\em bond} on the corresponding 
edge on $\mathcal{S}$, and to each extending edge not covered by a dimer, leave the corresponding 
edge {\em empty}. Then, as shown in Fig.~\ref{fig:vertex-dimer}, dimer coverings of a 
unit cell are mapped into vertex configurations on $\mathcal{S}$. Since the number of 
bonds extending from each vertex is either 1 or 3, which is an odd number, 
we are led to the {\em odd 8-vertex model} considered in \cite{fywu:jstatphys}.

Vertex weights of the odd 8-vertex model can be read off from Fig.~\ref{fig:vertex-dimer} as 
\be
\begin{array}{llll}
u_1=xz, & u_2=y, & u_3=y, & u_4=xz,\\
u_5=xy, & u_6=z, & u_7=z, & u_8=xy.
\end{array}
\label{equ:vertexweights}
\ee

The per-vertex 8-vertex model free energy  is then
\be
f_{\mathrm{8v}}(x,y,z)=\lim_{\siteno\rightarrow \infty}{\frac{1}{\siteno/3}
\ln Z_{\mathrm{kagome}}(x,y,z)} \ .
\label{eqn:8vfree}
\ee
Comparing Eq.~(\ref{eqn:8vfree}) with Eq.~(\ref{equ:deff}), we obtain the equivalence
\be
 f_{\mathrm{kagome}}(x,y,z)\ 
=\left (\frac{2}{3}\right )f_{\mathrm{8v}}(x,y,z)\ .
\label{equ:fkagomeasf8v}
\ee

Now the weights (\ref{equ:vertexweights}) satisfy the free-fermion condition
\be
u_1 u_2+u_3 u_4=u_5 u_6+u_7 u_8
\label{equ:freefermion}
\ee
for which the per-vertex 8-vertex model free energy has been evaluated in \cite{fywu:jstatphys} as 
\be
f_{\mathrm{8v}}=\frac{1}{16\pi^2}\int_{0}^{2\pi}{ \dif\theta \int_{0}^{2\pi}{ \dif\phi \ln F(\theta,\phi) } }
\label{equ:f8vasintF}
\ee
where 
\be
\begin{split}
F(\theta,\phi)=&2A+2D\cos(\theta-\phi)+2E\cos(\theta+\phi)\\
&+4\Delta_1 \sin^2\phi+4\Delta_2 \sin^2\theta
\nonumber
\end{split}
\ee
with
\be
\begin{array}{rcl}
A_{\phantom{2}} & = & \phantom{-}(u_1 u_3+u_2 u_4)^2+(u_5 u_7+u_6 u_8)^2\\
D_{\phantom{2}} & = & \phantom{-}(u_5 u_7)^2+(u_6 u_8)^2-2 u_1 u_2 u_3 u_4\\
E_{\phantom{2}} & = & -(u_1 u_3)^2-(u_2 u_4)^2+2 u_5 u_6 u_7 u_8\\
\Delta_1 & = & \phantom{-}(u_1 u_2-u_5 u_6)^2\\
\Delta_2 & = & \phantom{-}(u_3 u_4-u_5 u_6)^2
\end{array}
\label{equ:ADEDel1Del2}
\ee
The solution (\ref{equ:result1}) is now obtained by substituting Eqs.~(\ref{equ:ADEDel1Del2}) into 
Eqs.~(\ref{equ:f8vasintF}) and (\ref{equ:fkagomeasf8v}).

\section{Dimer-dimer correlation}\label{sec:correlation} 
The dimer-dimer correlation function 
can be evaluated by 
either considering a perturbation of the  Pfaffian
as in \cite{montroll:JMP,fisher2:PR} or by applying the Grassmannian method of 
\cite{fendley:PRB,samuel:}, details of both approaches will be given elsewhere.
Here we sketch steps in the Pfaffian computation. 
  
Define for the lattice edge
connecting sites $i$ and $j$ in unit cell at ${\bf r} = (r_x, r_y)$
an edge {\it vacancy} number
\begin{eqnarray}
n_{ij,{\bf r}} &=& 1, \quad {\rm if\>\> {\it ij}\>\>is \>\>empty},
                   \nonumber \\       &=& 0, \quad {\rm if\>\> {\it ij}\>\>is \>\>occupied},
\end{eqnarray}
where  $\langle \cdot\rangle $ denotes
the configurational average. Then, the
correlation function between two dimers on edges $ij$ in  cell ${\bf r}_1$
and  $k\ell$ in  cell  ${\bf r}_2$ is
\begin{eqnarray}
c\,(ij, {\bf r}_1;k\ell, {\bf r}_2)  =
\langle {\bar n}_{ij, {\bf r}_1}\, {\bar n}_{k\ell, {\bf r}_1} \rangle 
- \langle {\bar n}_{ij; {\bf r}_1}\rangle \langle {\bar n}_{k\ell; {\bf r}_2}\rangle .\label{equ:correlation}
\end{eqnarray}
 
To make use of Eqs.~(\ref{equ:correlation})
we need to compute the dimer generating function with specific edge(s) missing. 
Let  $A$ be the antisymmetric matrix derived from the Kasteleyn orientation,
and let $A'$ denote the antisymmetric matrix derived from $A$ with edge $ij$, say
in computing $\l {\bar n}_{ij} \r$, missing.
Write 
\begin{eqnarray}
Z&=& {\rm Pf} A \nonumber \\
Z' &=&  {\rm Pf} A' = {\rm Pf} [A+\Delta]
\end{eqnarray}
where $\Delta$ is the matrix with zero elements everywhere
except the $ij$ element is $-A_{ij}$ and the $ji$ element is $-A_{ji}\,(=A_{ij})$.
Then
\be
\l {\bar n}_{ij} \r = {Z'}/ Z = {\rm Pf} A' / {\rm Pf} A
\ee
and
\be
\l {\bar n}_{ij} \r ^2 = \frac {\det A'} {\det A} = \frac{ {\rm det }[ A (I + G \Delta)]} {\det A}
    = \det (I+G \Delta ).  \label{gf}
\ee
 where $ G = A^{-1}$ is the Greens function matrix and $I$ the identity matrix.

In computing Eq.~(\ref{gf}) we need only to keep those row(s) and column(s) in 
$\Delta$ and $A^{-1}$ where elements of $\Delta$ are nonzero. 
In addition, in the interior of a large lattice, 
the correlation depends only on the difference ${\bf r} = {\bf r_1} - {\bf r_2} = \{r_x, r_y\}$,
so elements of $G$ are given by
\begin{equation}
G({\bf r}) = \frac 1 {(2\pi)^2}
\int_0^{2\pi} \int_0^{2\pi} {\rm d} \t \,{\rm d}\p\ \,  e^{i(r_{x}\t +r_{y}\p)} A^{-1}(\t,\p). \label{inverse}
\end{equation}
These considerations lead to the explicit evaluations of Eq.~(\ref{gf}), and hence the correlation
(\ref{equ:correlation}).

Particularly, due to the fact that elements in $A^{-1}(\t, \p)$ contain only a monomial
of  $e^{\pm i\t}$ and $e^{\pm i\p}$, a consequence of
the fact that the determinant det$A$ is
given by the simple expression (\ref{Equ:determinant}), the 
integral (\ref{inverse}) vanishes  for 
$|r_{1x}-r_{2x}| > 1$ or $ |r_{1y}-r_{2y}| >1$. This leads to the result
 \be
c\,(ij, {\bf r}_1;k\ell, {\bf r}_2) = 0, \quad |\,{\bf r}_1 - {\bf r}_2| \geq 2.
\ee
The absence of the dimer-dimer correlation beyond a certain distance, which  is 
also found in 
the Sutherland-Rokhsar-Kivelson state of 
a quantum dimer model \cite{misguich:},
is a property unique to the kagome lattice. This underscores the special
role played by the kagome lattice in the statistical mechanics and quantum physics
of lattice systems.
 
{\it Acknowledgment}:  
One of us (FW) is supported in part by DOE grant LDRD DEA 3664LV.


\begin{thebibliography}{}
\bibitem[16]{boundary:} This argument holds as long as the transition cycle does not loop around the lattice.
\end{thebibliography}
\end{document}